\documentclass[conf]{new-aiaa}
\usepackage[utf8]{inputenc}
\usepackage{subdepth}
\usepackage{graphicx}
\usepackage{verbatim}
\usepackage{amsmath}
\usepackage[version=4]{mhchem}
\usepackage{siunitx}
\usepackage{longtable,tabularx}
\usepackage{multicol}
\usepackage{subcaption} 
\usepackage{algorithm}
\usepackage{algpseudocode}
\usepackage{todonotes}

\title{A comparative study of uncertainty quantification methods in gust response analysis of a Lift-Plus-Cruise eVTOL aircraft wing}

\author{Bingran Wang\footnote{Postdoctoral Scholar, Department of Mechanical and Aerospace Engineering, AIAA Member.}, 
Michael Warner\footnote{Ph.D Student, Department of Mechanical and Aerospace Engineering, AIAA Student Member.},
Aoran Tian\footnote{M.S. Student, Department of Mechanical and Aerospace Engineering, AIAA Student Member.},
Luca Scotzniovsky\footnote{Ph.D Student, Department of Mechanical and Aerospace Engineering, AIAA Student Member.}
and John T. Hwang\footnote{Associate Professor, Department of Mechanical and Aerospace Engineering, AIAA Member.}}
\affil{University of California, San Diego, La Jolla, CA, 92093}

\begin{document}

\maketitle
 \begin{abstract}
Wind gusts, being inherently stochastic, can significantly influence the safety and performance of aircraft. This study investigates a three-dimensional uncertainty quantification (UQ) problem to explore how uncertainties in gust and flight conditions affect the structural response of a Lift-Plus-Cruise eVTOL aircraft wing. The analysis employs an unsteady aeroelastic model with a one-way coupling between a panel method aerodynamic solver and a shell analysis structural solver to predict the wing's response under varying conditions. Additionally, this paper presents a comparative evaluation of commonly used non-intrusive UQ methods, including non-intrusive polynomial chaos, kriging, Monte Carlo, univariate dimension reduction, and gradient-enhanced univariate dimension reduction. These methods are assessed based on their effectiveness in estimating various risk measures—mean, standard deviation, and 95th percentile—of critical structural response outputs such as maximum tip displacement and average strain energy.
The numerical results reveal significant variability in the structural response outputs, even under relatively small ranges of uncertain inputs. This highlights the sensitivity of the system to uncertainties in gust and flight conditions. Furthermore, the performance of the implemented UQ methods varies significantly depending on the specific risk measures and the quantity of interest being analyzed.
\end{abstract}
\section{Introduction}
The accurate prediction of gust loads is critical for aircraft design and certification, as gust-induced aeroelastic responses can significantly impact the safety and performance of an aircraft. Gust loads arise from atmospheric turbulence and are particularly challenging to model due to their dynamic and stochastic nature. These loads must be carefully analyzed to ensure compliance with regulatory requirements and to maintain the safety margins necessary for aircraft certification. Traditional deterministic approaches often fail to capture the inherent uncertainties in gust characteristics and flight conditions. This underscores the importance of incorporating uncertainty quantification (UQ) into aeroelastic analysis to better understand and predict the variability in structural responses~\cite{beran2017uncertainty}.

UQ is a crucial tool in engineering, which enables engineers to evaluate how input uncertainties influence the system’s outputs.
Its applications span diverse fields, including weather forecasting~\cite{joslyn2010communicating,hess-9-381-2005}, machine learning~\cite{hullermeier2021aleatoric}, structural analysis~\cite{wan2014analytical,hu2018uncertainty}, and aircraft design~\cite{ng2016monte,wang2024graph,lim2022uncertainty}.
Common non-intrusive UQ methods include polynomial chaos, kriging, and Monte Carlo. Kriging, also known as Gaussian process regression, constructs a surrogate response surface from input-output data, enabling efficient model evaluations for reliability analysis~\cite{kaymaz2005application,hu2016single}. The polynomial chaos method represents quantities of interest (QoIs) using orthogonal polynomials tailored to the input distributions, leveraging smoothness in the random space for rapid convergence with either a sampling or an integration approach~\cite{hosder2006non,jones2013nonlinear,keshavarzzadeh2017topology}. Both kriging and polynomial chaos are effective for low- to medium-dimensional problems but face scalability challenges as dimensionality increases, limiting their direct application in high-dimensional UQ problems. Conversely, the Monte Carlo method uses random sampling to estimate the risk measures of the QoIs. This method is not only easy to implement but also well-suited for high-dimensional problems due to their convergence rate being independent of input dimensionality.
Several approximation methods are also widely used to reduce computational costs without significant loss of accuracy. These include active subspace ~\cite{constantine2014active}, first-order Taylor expansion~\cite{wooldridge2001applications}, univariate dimension reduction (UDR)~\cite{constantine2014active}, and gradient-enhanced dimension reduction~\cite{wang2024gudr} methods. Such methods aim to strike a balance between computational efficiency and predictive accuracy by reducing the problem dimension or creating an approximation function for the model, making them practical for large-scale or high-fidelity engineering problems. 
Each UQ method has its strengths and limitations, with performance varying depending on the specific problem setup and characteristics. Consequently, selecting the most appropriate method requires careful consideration of the problem's dimensionality, complexity, and desired accuracy.

In previous work on aircraft gust response analysis, Pettit et al.~\cite{pettit2007gust} conducted UQ on an aerodynamic model with uncertain gust velocity spectra using the NIPC method. Similarly, Cook et al.~\cite{cook2018efficient} applied UQ to aeroelastic analysis, accounting for structural and flight condition uncertainties, also leveraging NIPC. Xiang's thesis~\cite{xiang2024time} extended this research by performing time-domain structural optimization of eVTOL wings under gust loads. However, UQ problems involving unsteady aeroelastic analysis remain underexplored, particularly when considering uncertainties in gust conditions.
This work focuses on a Lift-Plus-Cruise (LPC) eVTOL aircraft design. The development of eVTOL aircraft has gained significant attention in academia due to their potential to transform next-generation transportation systems. 
Recent studies on eVTOL design have emphasized multidisciplinary design optimization~\cite{sarojini2023large}, trajectory optimization~\cite{chauhan2020tilt}, and acoustic modeling~\cite{li2024analytic}. Despite these advancements, the impact of uncertain gust and flight conditions on eVTOL performance remains an open question.

In this paper, we formulate a three-dimensional UQ problem that considers uncertainties in gust and flight conditions. An unsteady aeroelastic simulation model is used to predict the structural response of the wing under varying conditions. Specifically, the aeroelastic simulation integrates a mid-fidelity panel method aerodynamic solver with a high-fidelity shell-based structural solver using a one-way coupling approach. 
The computational model is implemented on a graph-based modeling framework, \textit{Computational System Design
Language} (CSDL)~\cite{gandarillas2022novel}.
CSDL enables automatic differentiation~\cite{sperry2023automatic} and computational graph transformation~\cite{wang2023accelerating, wang2024partial} to reduce the computational cost for UQ and optimization.
The QoIs in this UQ problem are the maximum tip displacement and average strain energy.
To estimate the three risk measures—mean, standard deviation, and 95th percentile—of the QoIs, we implement five UQ methods: non-intrusive polynomial chaos, kriging, Monte Carlo, univariate dimension reduction, and gradient-enhanced univariate dimension reduction. 

The numerical results reveal significant variability in both QoIs—maximum tip displacement and average strain energy—even under small variations in the uncertain gust and flight conditions. This highlights the sensitivity of the structural response to these uncertainties and underscores the importance of incorporating UQ in gust response analysis for aircraft design.

Among the UQ methods compared, kriging demonstrated superior performance in estimating risk measures for maximum tip displacement, as its interpolation-based approach effectively handles highly nonlinear response surfaces. In contrast, NIPC excelled in estimating risk measures for average strain energy, as its underlying function exhibits smoother behaviour.
In the meantime, UDR and GUDR are cheap alternative methods for estimating various risk measures, while Monte Calro underperformed due to the low dimensionality of the UQ problem.

This paper is organized as follows. 
Section \ref{Sec: Problem} presents detailed description on the aeroelastic simulation model and the UQ problem. 
Section \ref{sec: results} shows the numerical results of the implemented UQ methods.
Section \ref{Sec: Conclusion} summarizes the work and offers concluding thoughts.


\section{Problem set-up}
\label{Sec: Problem}

\begin{figure}
    \centering
    \includegraphics[width=1\linewidth]{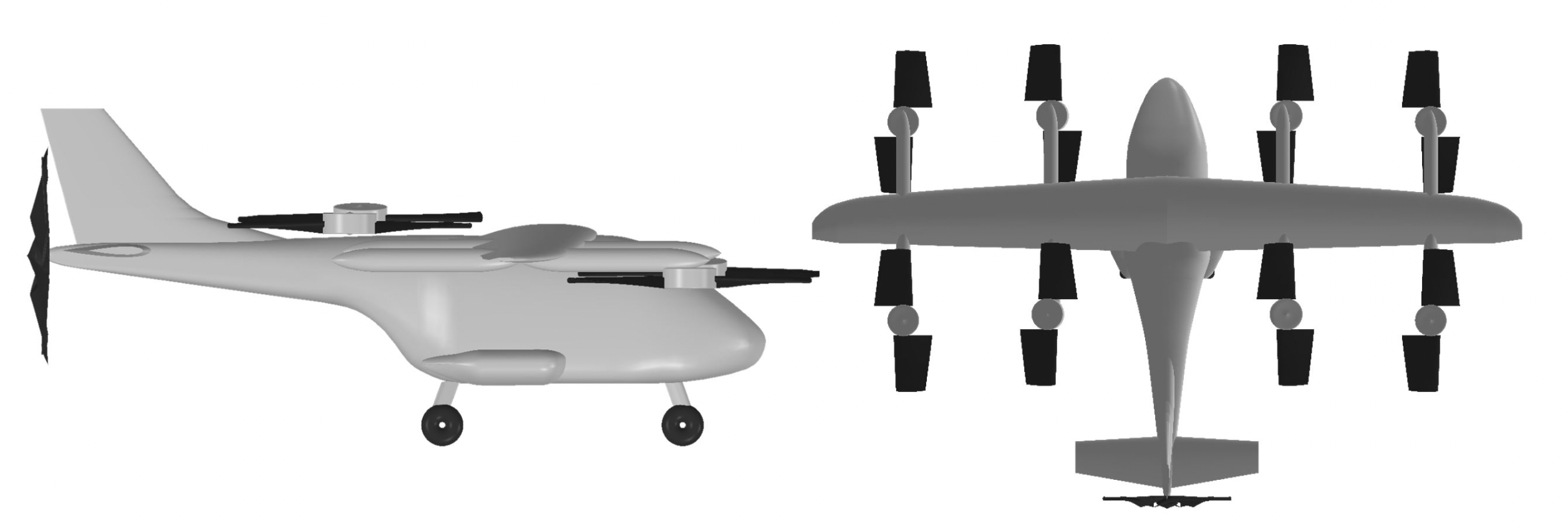}
    \caption{Lift-Plus-Cruise eVTOL aircraft concept}
    \label{fig:lift_plus_cruise}
\end{figure}

In this paper, we address an uncertainty quantification problem related to the structural response of a Lift-Plus-Cruise eVTOL aircraft wing under stochastic flight and gust conditions. The Lift-Plus-Cruise (LPC) concept is a well-established design for eVTOL aircraft, characterized by dedicated rotors for vertical takeoff and landing, as well as fixed wings and a rear pusher propeller for cruise flight. The Lift-Plus-Cruise concept is illustrated in Fig.~\ref{fig:lift_plus_cruise}. To investigate the structural response of the wing under different conditions, we conduct an unsteady aeroelastic analysis using a high-fidelity structural solver based on shell analysis, coupled with a mid-fidelity aerodynamic solver employing the panel method. 

\subsection{Unsteady aeroeleastic simulation}
\begin{figure}
    \centering
    \includegraphics[width=0.5\linewidth]{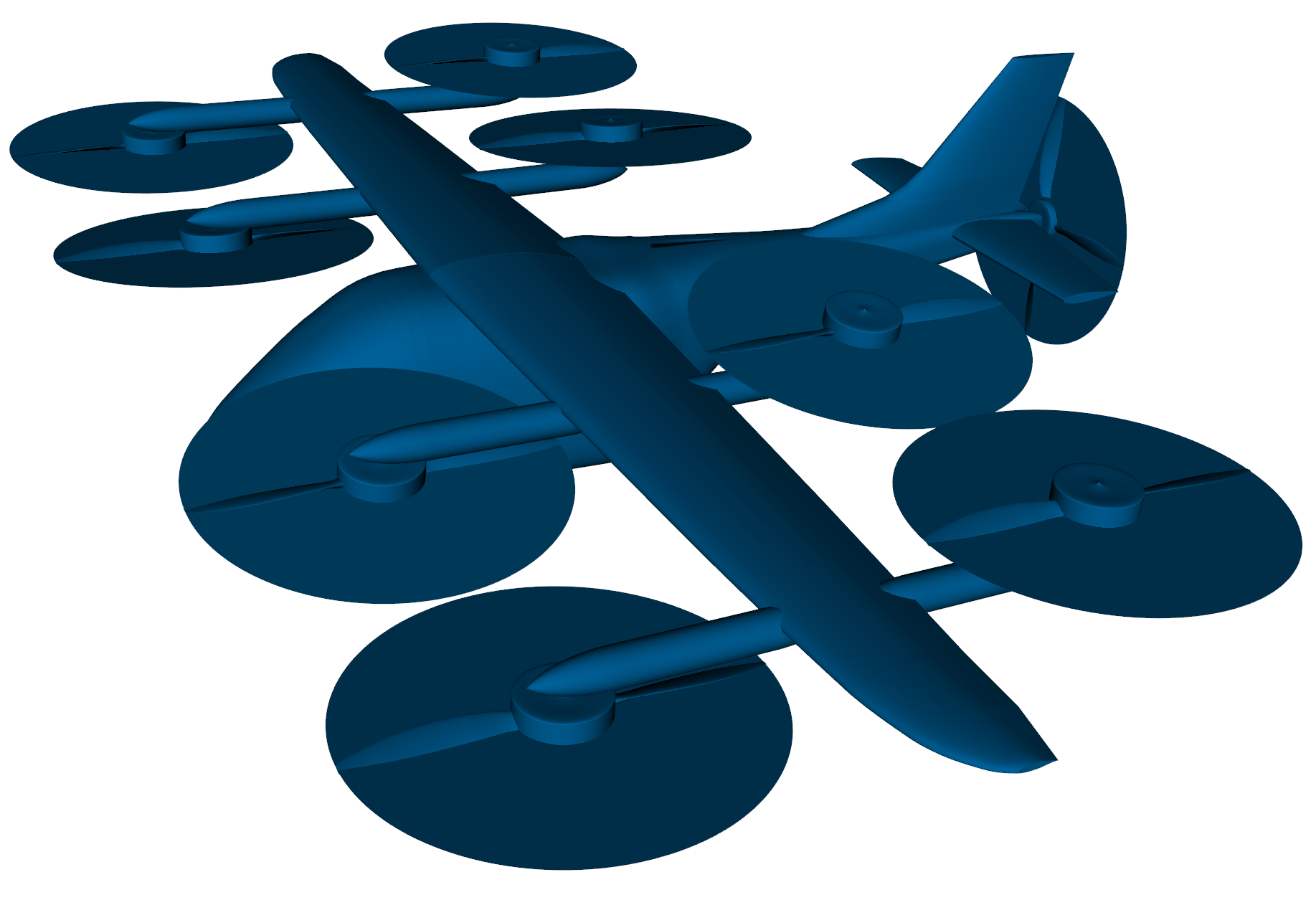}
    \caption{Lift-Plus-Cruise eVTOL aircraft geometry}
    \label{fig:lpc_geo}
\end{figure}

\begin{figure}[ht!]
    \centering
    \begin{subfigure}[t]{\textwidth}
        \centering
        \includegraphics[width=1\textwidth]{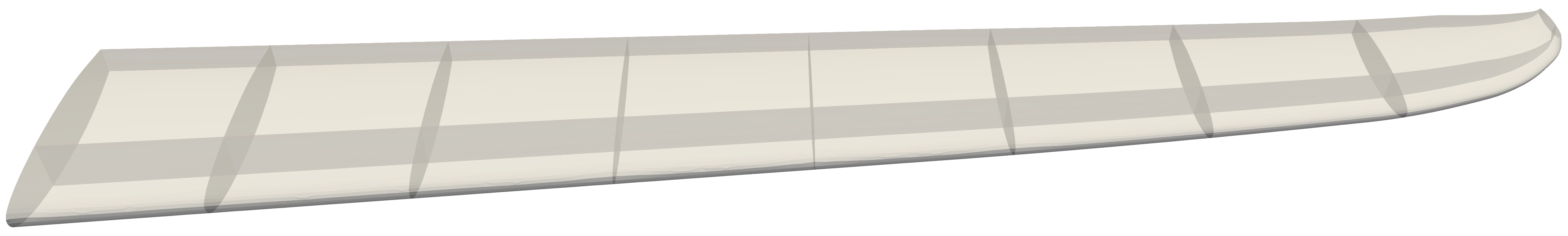} 
        \caption{Internal structures of the LPC aircraft wing}
        \label{fig:top}
    \end{subfigure}
    \vspace{1cm}
    \begin{subfigure}[t]{\textwidth}
        \centering
        \includegraphics[width=1\textwidth]{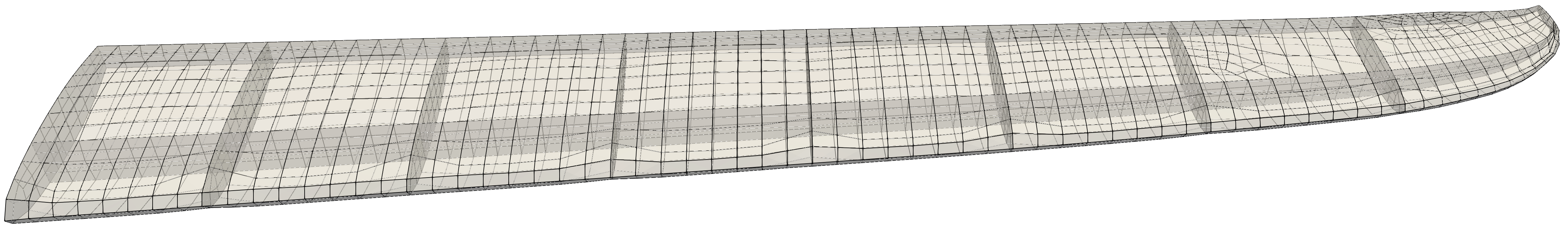} 
        \caption{Shell mesh used for structural analysis}
        \label{fig:bottom}
    \end{subfigure}

    \caption{LPC aircraft wing structure and the shell mesh used for structural analysis}
    \label{fig:int_struc}
\end{figure}

For the aeroelastic simulation, we consider a full-scale Lift-Plus-Cruise (LPC) eVTOL aircraft, with its geometry depicted in Fig.~\ref{fig:lpc_geo}. The simulation model is adapted from the work in~\cite{xiang2024time}. This study focuses on the left wing of the LPC aircraft, which features internal structures consisting of 9 ribs and 2 spars. These internal structural components are illustrated in Fig.~\ref{fig:int_struc}.

Aeroelastic simulation involves the coupled analysis of aerodynamic forces and structural dynamics to predict the behavior of flexible aircraft structures under various conditions, such as gust excitations. In this work, the structural solver is implemented based on Reissner-Mindlin shell theory, which accounts for transverse shear deformation and rotary inertia. This makes it particularly suitable for analyzing lightweight, flexible wings typical of eVTOL designs. A relatively coarse shell mesh with around 2000 cells is used for shell analysis, the shell mesh is depicted in Fig.~\ref{fig:int_struc}.
We use the unsteady panel method, an aerodynamic analysis method based on potential flow theory, to model aerodynamics \cite{katz2001low, epton1990pan, hess1967calculation, maskew1987program}. The panel method computes the surface pressure field and aerodynamic forces based on the oncoming free-stream flow and the flow perturbation due to gust. The unsteady wake effects are also captured, which ensures that the accurate aerodynamic loads are captured from gust excitations. 

To reduce the computational cost of the unsteady aeroelastic simulation, the aerodynamic and structural solvers are connected using a one-way coupling approach. In this framework, the aerodynamic pressure profile is computed first using an unsteady panel method. These aerodynamic loads are then passed to the structural shell analysis solver, which calculates the structural response, including displacements, internal stresses, and strain energy. This one-way coupling simplifies the simulation process by avoiding iterative feedback between the solvers, making it computationally efficient while still capturing critical aeroelastic dynamics.

\subsection{Gust modeling}
\begin{figure}
    \centering
    \includegraphics[width=1\linewidth]{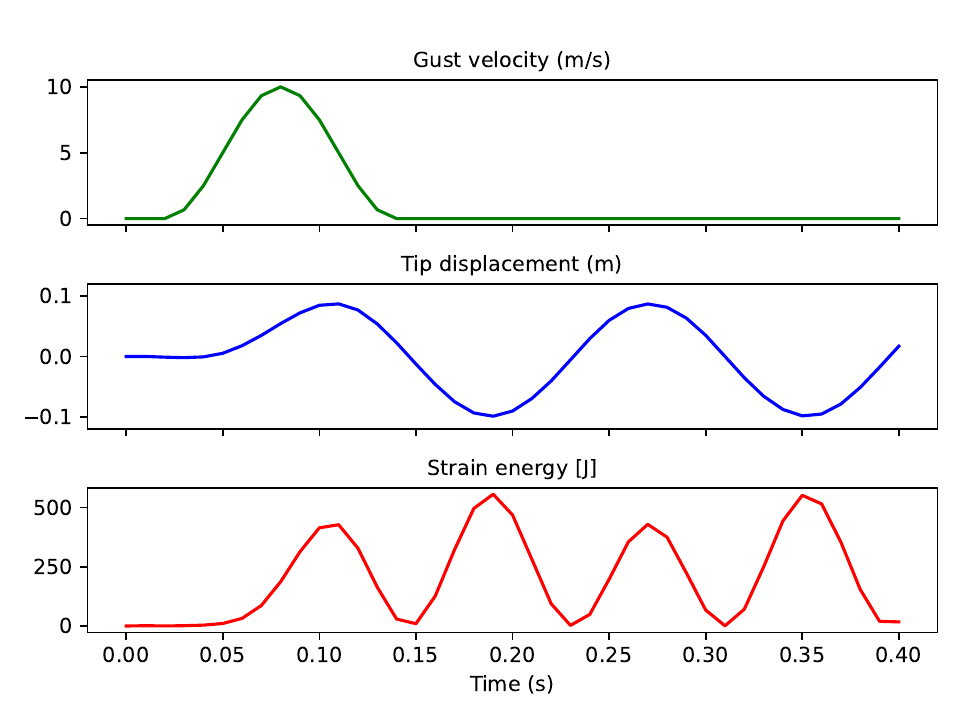}
    \caption{Gust profile and the numerical results of the aeroelastic simulation}
    \label{fig:gust_response}
\end{figure}

Gust modeling is a crucial aspect of aeroelastic analysis, especially for assessing the response of flexible aircraft structures to atmospheric disturbances. 
In this study, we employ a discrete gust model to simplify the atmospheric turbulence into a well-defined velocity profile.
Specifically, we use a one-minus-cosine gust profile which features a smooth, sinusoidal-like variation in vertical velocity.   
This approach is widely used in aeroelastic simulations as it balances accuracy and feasibility when integrated with finite element models and aeroelastic solvers.
For the dynamic scenario considered, the eVTOL wing is subjected to a vertical one-minus-cosine gust while maintaining steady horizontal flight with a freestream velocity of \(V_\infty\). The vertical gust velocity, \(V_g(t)\), is defined as:
\begin{equation}
    V_g(t) =
\begin{cases} 
\frac{1}{2}V_p \left(1 - \cos\left(\frac{2\pi(t - T_0)V_\infty}{l_g}\right)\right) & T_0 < t < T_0 + l_g/V_\infty, \\
0 & \text{otherwise},
\end{cases}
\end{equation}
where:
\begin{itemize}
    \item \(l_g\): gust length,
    \item \(V_p\): peak gust velocity,
    \item \(T_0\): time of gust onset.
\end{itemize}
This profile induces a transient aerodynamic loading that drives the structural response of the wing. Fig.~\ref{fig:gust_response} illustrates the gust velocity profile along with numerical results from the aeroelastic simulation, where the time step is set to \(\Delta t = 0.01 \, \text{s}\).
From the numerical results, we observe that the maximum tip displacement occurs shortly after the gust velocity reaches its peak. This behavior is expected, as the wing tip continues its upward motion due to its nonzero velocity at the moment of peak gust velocity. The upward movement persists until the tip velocity reduces to zero. Furthermore, the oscillation amplitude of the tip displacement remains constant because the elastic model does not incorporate damping effects. This highlights the importance of considering damping in future analyses for more realistic predictions of structural behavior.

\subsection{Uncertainty quantification problem}

In this study, we formulate a 3-dimensional UQ problem to analyze the impact of uncertain input parameters on the aeroelastic response of a LPC eVTOL aircraft wing under gust excitations. 
We identified three key uncertain inputs, as summarized in Table~\ref{tab:uncertain_inputs}: flight velocity ($V_{\infty}$), gust length ($l_g$), and gust peak velocity ($V_p$). Each input is modeled as an independent random variable with a uniform probability distribution to reflect variability in operating and environmental conditions. The ranges of these distributions are chosen based on typical operational scenarios for eVTOL aircraft and typical gust conditions.

\begin{table}[h!]
    \centering
    \caption{Uncertain inputs and their distributions}\label{tab:uncertain_inputs}
    \begin{tabular}{c | c }
         \textbf{Uncertain Inputs} & \textbf{Distributions} \\
         \hline
         Flight velocity $V_{\infty}$ ($m/s$) & $\mathcal{U}(40, 60)$ \\
         Gust length $l_g$ ($m$) & $\mathcal{U}(4, 8)$ \\
         Gust peak velocity $V_p$ ($m/s$) & $\mathcal{U}(5, 15)$ \\
    \end{tabular}
\end{table}

The uncertain flight velocity ($V_{\infty}$) reflects variability in the operating speeds of an eVTOL aircraft. The gust length ($l_g$) represents the spatial extent of the gust disturbance, which may vary depending on atmospheric conditions. The gust peak velocity ($V_p$) accounts for the intensity of the gust and is critical for assessing structural loads during transient events.

The UQ problem aims to propagate these uncertainties through the aeroelastic simulation model to quantify their effects on key output QoIs. The QoIs considered in this study include maximum tip displacement and the average strain energy, which are critical indicators of structural performance under gust loads. Risk measures such as mean, standard deviation, and the 95th percentile are computed for each QoI to provide a comprehensive assessment of the system's response variability.
\section{Numerical Results}
\label{sec: results}
\begin{table}[h!]
    \centering
    \caption{Ground truth UQ results}\label{tab:uncertain_results}
    \begin{tabular}{c | c c c}
         \textbf{Quantity of interests} & \textbf{mean} & \textbf{standard deviation} & \textbf{95th percentile} \\
         \hline
         Maximum displacement (m) & 0.0843 & 0.0179 & 0.111 \\
         Average strain energy (J) & 208 & 60.9 & 313\\
    \end{tabular}
\end{table}
\begin{figure}
    \centering
    \includegraphics[width=1\linewidth]{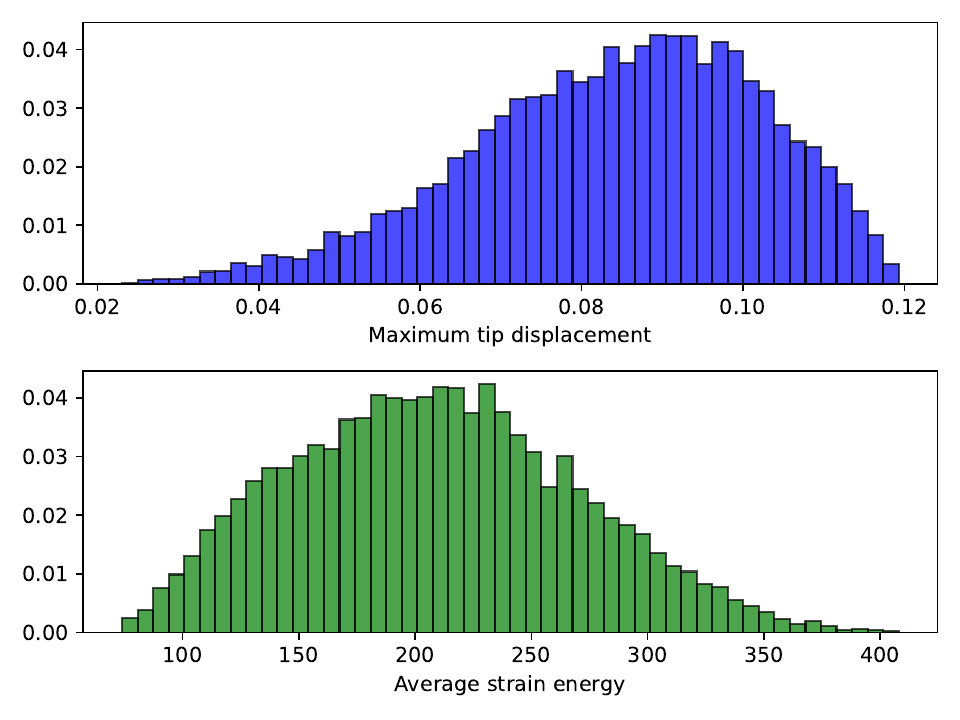}
    \caption{Probability density distributions of the QoIs (maximum tip displacement and average strain energy)}
    \label{fig:pdf_results}
\end{figure}
In this study, we implemented five uncertainty quantification (UQ) methods to evaluate their performance on the formulated UQ problem. The methods implemented are as follows:
\begin{itemize}
    \item \textbf{Non-Intrusive Polynomial Chaos (NIPC)}: A non-adaptive regression-based NIPC method is implemented using Chaospy~\cite{feinberg2015chaospy}.
    \item \textbf{Kriging}: a non-adaptive kriging method is implemented using the surrogate modelling toolbox~\cite{SMT2019}.
    \item \textbf{Monte Carlo}: a brute force non-adaptive Monte Carlo method is implemented.
    \item \textbf{Univariate dimension reduction (UDR)}: the UDR method is implemented following~\cite{rahman2004univariate}; the 95\% quantile is estimated using a constructed polynomial chaos surrogate of the UDR approximation function.
    \item \textbf{Gradient-enhanced univariate dimension reduction}: the GUDR method is implemented following~\cite{wang2024gudr}; the 95\% quantile is estimated using a constructed polynomial chaos surrogate of the GUDR approximation function. The gradients are computed through automatic differentiation in CSDL~\cite{gandarillas2022novel}.
\end{itemize}
In this study, we applied five UQ methods to compute three risk measures—mean, standard deviation, and 95th percentile—of two QoIs: maximum tip displacement and average strain energy. The ground truth results were estimated using the kriging method with 500 sample points, providing a highly accurate benchmark for comparison. The ground truth UQ results are summarized in Table~\ref{tab:uncertain_results}, and the probability density functions (PDFs) of the QoIs are plotted in Fig.~\ref{fig:pdf_results}.
The results indicate that both QoIs exhibit significant variability, even though the ranges of the uncertain inputs are relatively small. Additionally, the PDFs in Fig.~\ref{fig:pdf_results} reveal that the distribution of average strain energy has a pronounced right tail, making the estimation of its 95th percentile more challenging compared to maximum tip displacement, which exhibits a left-tailed distribution.
\begin{figure}
    \centering
    \includegraphics[width=1\linewidth]{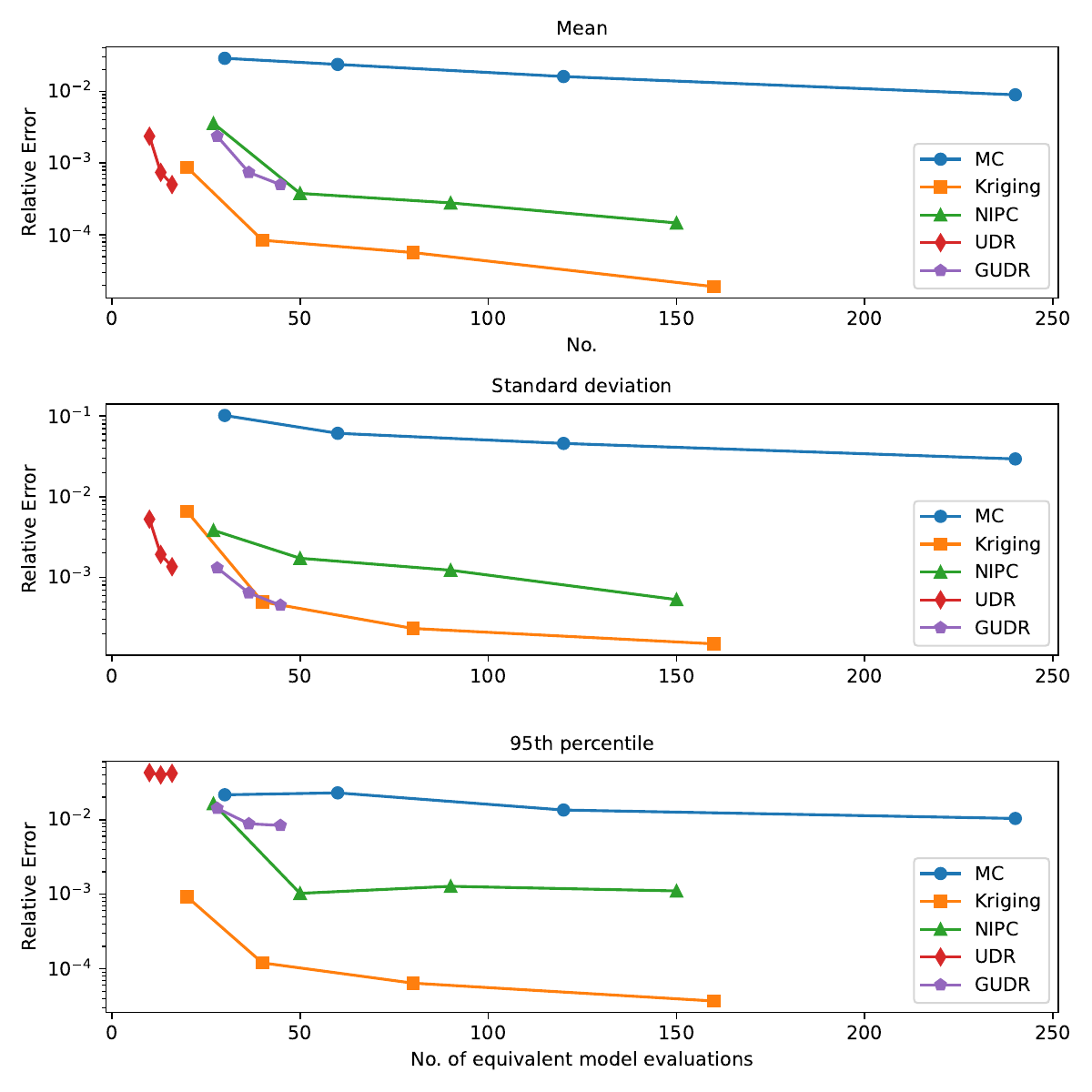}
    \caption{UQ convergence plots for maximum tip displacement}
    \label{fig:uq_result_max_disp}
\end{figure}

\begin{figure}
    \centering
    \includegraphics[width=1\linewidth]{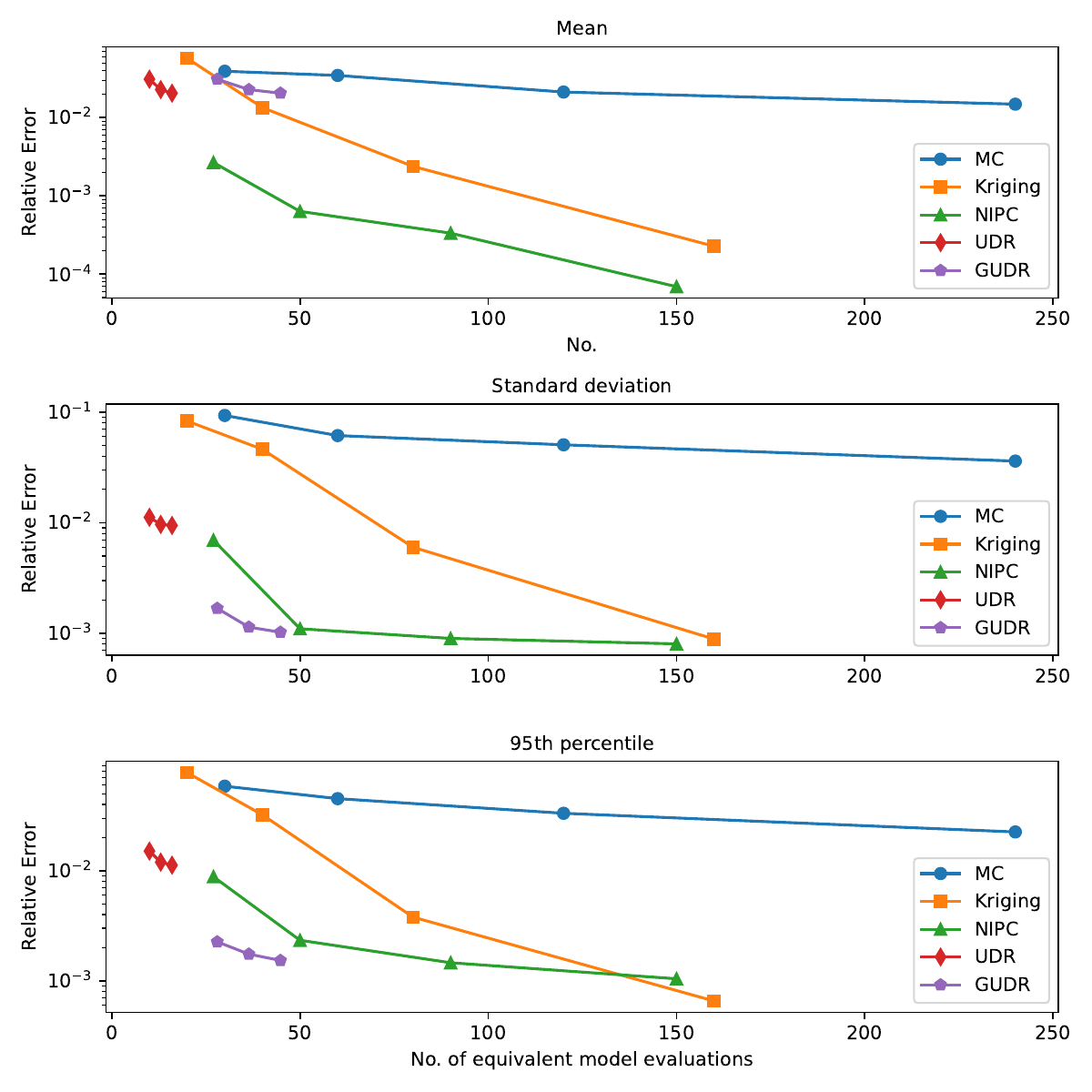}
    \caption{UQ convergence plots for average strain energy}
    \label{fig:uq_result_strain}
\end{figure}

The convergence plots of each UQ method of interest, for maximum tip displacement and average strain energy, are presented in Fig.~\ref{fig:uq_result_max_disp} and ~\ref{fig:uq_result_strain}, respectively.
From the results, it is evident that all of the implemented UQ methods performed reasonably well on this problem, achieving relative errors below $1e-2$ in most cases. However, the performance of the methods varied depending on the specific risk measure and QoI being analyzed.

The Monte Carlo method consistently underperformed compared to the other methods due to the low dimensionality of the UQ problem, which involves only three uncertain inputs. The other UQ methods demonstrated varying degrees of effectiveness depending on the characteristics of the QoIs. Specifically, kriging outperformed NIPC in estimating the risk measures for maximum tip displacement, while NIPC excelled in estimating those for average strain energy. This difference is attributed to the underlying nature of the specific outputs. The function to compute the maximum tip displacement involves a \textit{max} operation that introduces high nonlinearity, favouring interpolation-based methods like kriging. In contrast, the NIPC method is favored for the average strain energy as it exhibits smoother behavior, making it easier to approximate with polynomials.

The characteristics of the QoIs also influenced the performance of UDR and GUDR, which rely on quadrature-based approaches for estimating risk measures. 
As a result, the UDR and GUDR methods behaved better than kriging in estimating the risk of measures of the average strain energy but were outperformed in estimating the risk measures of the maximum tip displacement. 
Comparing UDR with GUDR, we see similar results estimating the mean using these two methods, with UDR being computationally less expensive. This is consistent with the theory in~\cite{wang2024gudr}. 
However, the accuracy of UDR decreases when estimating higher-order risk measures such as standard deviation and 95th percentile. The accuracy of these higher-order metrics is improved by using GUDR because the univariate gradient information is incorporated. By utilizing the automatic differentiation capabilities of CSDL, the gradients used for GUDR were efficiently computed; as a result, this made GUDR highly effective in estimating the standard deviation and 95th percentile values of the average strain energy.
When comparing risk measures, mean values were consistently estimated with the highest accuracy across all methods, while the 95th percentile proved to be the most challenging to estimate accurately. This highlights the importance of employing adaptive UQ methods like importance sampling when estimating reliability measures. Additionally, when estimating the 95th percentile, the performance of all UQ methods on maximum tip displacement is better than average strain energy. This is because the right-trailed PDF of average strain energy (shown in Fig.~\ref{fig:pdf_results}) makes the 95th percentile harder to estimate.

Overall, for low-dimensional UQ problems, kriging and NIPC are effective choices, while UDR and GUDR serve as low-cost alternatives. For functions with strong nonlinearity or discontinuities, interpolation-based methods like kriging are more effective. Conversely, for functions that can be efficiently approximated with polynomials, NIPC is preferable. UDR provides reasonably accurate results with minimal computational cost, while GUDR enhances accuracy for higher statistical moments and reliability measures when efficient gradient computation is available.
\section{Conclusion}
\label{Sec: Conclusion}
In this paper, we present a comprehensive uncertainty analysis for the gust response of the wing of a Lift-Plus-Cruise eVTOL aircraft to investigate the impact of uncertainties in gust and flight conditions on structural performance. A time-dependent aeroelastic simulation framework was utilized, integrating a vortex lattice method aerodynamic solver with a shell-based structural solver with one-way coupling. The QoIs—maximum tip displacement and average strain energy—were analyzed to compute three risk measures: mean, standard deviation, and 95th percentile.

Five UQ methods were implemented and compared: non-intrusive polynomial chaos, kriging, Monte Carlo, univariate dimension reduction (UDR), and gradient-enhanced univariate dimension reduction (GUDR).
Numerical results show significant variability of both QoIs under small variations of the uncertain gust and flight conditions.
Comparing the UQ methods, kriging demonstrated superior performance for estimating risk measures of maximum tip displacement due to its ability to handle highly nonlinear response surfaces.
In contrast, NIPC excelled in estimating risk measures for average strain energy, benefiting from the smoothness of the underlying function. 
UDR provided accurate and computationally efficient estimates for mean values but showed reduced accuracy for standard deviation and 95th percentile. GUDR improved upon UDR by incorporating gradient information, excelling in estimating higher-order moments and reliability measures when gradient evaluations can be efficiently performed.

This study highlights the importance of incorporating UQ analysis into aircraft design, particularly for gust response analysis, and also provides insights on how to choose the best UQ method based on the problem dimensionality, characteristics of the QoIs, and the desired accuracy for specific risk measures.
\bibliography{sample}

\begin{thebibliography}{34}
\newcommand{\enquote}[1]{``#1''}
\providecommand{\natexlab}[1]{#1}
\providecommand{\url}[1]{\texttt{#1}}
\providecommand{\urlprefix}{URL }
\expandafter\ifx\csname urlstyle\endcsname\relax
  \providecommand{\doi}[1]{\discretionary{}{}{}https://doi.org/#1}\else
  \providecommand{\doi}[1]{\discretionary{}{}{}\urlstyle{rm}\url{https://doi.org/#1}}\fi

\bibitem[{Beran et~al.(2017)Beran, Stanford, and Schrock}]{beran2017uncertainty}
Beran, P., Stanford, B., and Schrock, C., \enquote{Uncertainty quantification in aeroelasticity,} \emph{Annual review of fluid mechanics}, Vol.~49, No.~1, 2017, pp. 361--386.

\bibitem[{Joslyn and Savelli(2010)}]{joslyn2010communicating}
Joslyn, S., and Savelli, S., \enquote{Communicating forecast uncertainty: Public perception of weather forecast uncertainty,} \emph{Meteorological Applications}, Vol.~17, No.~2, 2010, pp. 180--195.
\newblock \doi{10.1002/met.190}.

\bibitem[{Pappenberger et~al.(2005)Pappenberger, Beven, Hunter, Bates, Gouweleeuw, Thielen, and de~Roo}]{hess-9-381-2005}
Pappenberger, F., Beven, K.~J., Hunter, N.~M., Bates, P.~D., Gouweleeuw, B.~T., Thielen, J., and de~Roo, A. P.~J., \enquote{Cascading model uncertainty from medium range weather forecasts (10 days) through a rainfall-runoff model to flood inundation predictions within the European Flood Forecasting System (EFFS),} \emph{Hydrology and Earth System Sciences}, Vol.~9, No.~4, 2005, pp. 381--393.
\newblock \doi{10.5194/hess-9-381-2005}.

\bibitem[{H{\"u}llermeier and Waegeman(2021)}]{hullermeier2021aleatoric}
H{\"u}llermeier, E., and Waegeman, W., \enquote{Aleatoric and epistemic uncertainty in machine learning: An introduction to concepts and methods,} \emph{Machine Learning}, Vol. 110, 2021, pp. 457--506.
\newblock \doi{10.1007/s10994-021-05946-3}.

\bibitem[{Wan et~al.(2014)Wan, Mao, Todd, and Ren}]{wan2014analytical}
Wan, H.-P., Mao, Z., Todd, M.~D., and Ren, W.-X., \enquote{Analytical uncertainty quantification for modal frequencies with structural parameter uncertainty using a Gaussian process metamodel,} \emph{Engineering Structures}, Vol.~75, 2014, pp. 577--589.
\newblock \doi{10.1016/j.engstruct.2014.06.028}.

\bibitem[{Hu et~al.(2018)Hu, Mahadevan, and Ao}]{hu2018uncertainty}
Hu, Z., Mahadevan, S., and Ao, D., \enquote{Uncertainty aggregation and reduction in structure--material performance prediction,} \emph{Computational Mechanics}, Vol.~61, No.~1, 2018, pp. 237--257.
\newblock \doi{10.1007/s00466-017-1448-6}.

\bibitem[{Ng and Willcox(2016)}]{ng2016monte}
Ng, L.~W., and Willcox, K.~E., \enquote{Monte Carlo information-reuse approach to aircraft conceptual design optimization under uncertainty,} \emph{Journal of Aircraft}, Vol.~53, No.~2, 2016, pp. 427--438.
\newblock \doi{10.2514/1.C033352}.

\bibitem[{Wang et~al.(2024{\natexlab{a}})Wang, Orndorff, Joshy, and Hwang}]{wang2024graph}
Wang, B., Orndorff, N.~C., Joshy, A.~J., and Hwang, J.~T., \enquote{Graph-accelerated large-scale multidisciplinary design optimization under uncertainty of a laser-beam-powered aircraft,} \emph{AIAA SCITECH 2024 Forum}, 2024{\natexlab{a}}, p. 0169.
\newblock \doi{10.2514/6.2024-0169}.

\bibitem[{Lim et~al.(2022)Lim, Kim, and Yee}]{lim2022uncertainty}
Lim, D., Kim, H., and Yee, K., \enquote{Uncertainty propagation in flight performance of multirotor with parametric and model uncertainties,} \emph{Aerospace Science and Technology}, Vol. 122, 2022, p. 107398.
\newblock \doi{10.1016/j.ast.2022.107398}.

\bibitem[{Kaymaz(2005)}]{kaymaz2005application}
Kaymaz, I., \enquote{Application of kriging method to structural reliability problems,} \emph{Structural safety}, Vol.~27, No.~2, 2005, pp. 133--151.
\newblock \doi{10.1016/j.strusafe.2004.09.001}.

\bibitem[{Hu and Mahadevan(2016)}]{hu2016single}
Hu, Z., and Mahadevan, S., \enquote{A single-loop kriging surrogate modeling for time-dependent reliability analysis,} \emph{Journal of Mechanical Design}, Vol. 138, No.~6, 2016, p. 061406.
\newblock \doi{10.1115/1.4033428}.

\bibitem[{Hosder et~al.(2006)Hosder, Walters, and Perez}]{hosder2006non}
Hosder, S., Walters, R., and Perez, R., \enquote{A non-intrusive polynomial chaos method for uncertainty propagation in CFD simulations,} \emph{44th AIAA aerospace sciences meeting and exhibit}, 2006, p. 891.
\newblock \doi{10.2514/6.2006-891}.

\bibitem[{Jones et~al.(2013)Jones, Doostan, and Born}]{jones2013nonlinear}
Jones, B.~A., Doostan, A., and Born, G.~H., \enquote{Nonlinear propagation of orbit uncertainty using non-intrusive polynomial chaos,} \emph{Journal of Guidance, Control, and Dynamics}, Vol.~36, No.~2, 2013, pp. 430--444.
\newblock \doi{10.2514/1.57599}.

\bibitem[{Keshavarzzadeh et~al.(2017)Keshavarzzadeh, Fernandez, and Tortorelli}]{keshavarzzadeh2017topology}
Keshavarzzadeh, V., Fernandez, F., and Tortorelli, D.~A., \enquote{Topology optimization under uncertainty via non-intrusive polynomial chaos expansion,} \emph{Computer Methods in Applied Mechanics and Engineering}, Vol. 318, 2017, pp. 120--147.
\newblock \doi{10.1016/j.cma.2017.01.019}.

\bibitem[{Constantine et~al.(2014)Constantine, Dow, and Wang}]{constantine2014active}
Constantine, P.~G., Dow, E., and Wang, Q., \enquote{Active subspace methods in theory and practice: applications to kriging surfaces,} \emph{SIAM Journal on Scientific Computing}, Vol.~36, No.~4, 2014, pp. A1500--A1524.
\newblock \doi{10.1137/130916138}.

\bibitem[{Wooldridge(2001)}]{wooldridge2001applications}
Wooldridge, J.~M., \enquote{Applications of generalized method of moments estimation,} \emph{Journal of Economic perspectives}, Vol.~15, No.~4, 2001, pp. 87--100.
\newblock \doi{10.1257/jep.15.4.87}.

\bibitem[{Wang et~al.(2024{\natexlab{b}})Wang, Orndorff, Sperry, and Hwang}]{wang2024gudr}
Wang, B., Orndorff, N.~C., Sperry, M., and Hwang, J.~T., \enquote{A gradient-enhanced univariate dimension reduction method for uncertainty propagation,} \emph{Aerospace Science and Technology}, Vol. 155, 2024{\natexlab{b}}, p. 109602.

\bibitem[{Pettit et~al.(2007)Pettit, Hajj, and Beran}]{pettit2007gust}
Pettit, C., Hajj, M., and Beran, P., \enquote{Gust loads with uncertainty due to imprecise gust velocity spectra,} \emph{48th AIAA/ASME/ASCE/AHS/ASC Structures, Structural Dynamics, and Materials Conference}, 2007, p. 1965.

\bibitem[{Cook et~al.(2018)Cook, Wales, Gaitonde, Jones, and Cooper}]{cook2018efficient}
Cook, R.~G., Wales, C., Gaitonde, A., Jones, D., and Cooper, J.~E., \enquote{Efficient modelling of a nonlinear gust loads process for uncertainty quantification of highly flexible aircraft,} \emph{2018 AIAA/ASCE/AHS/ASC Structures, Structural Dynamics, and Materials Conference}, 2018, p. 1681.

\bibitem[{Xiang(2024)}]{xiang2024time}
Xiang, R., \enquote{Time-domain structural optimization of eVTOL wings under gust loads using automated adjoint methods,} Ph.D. thesis, University of California, San Diego, 2024.

\bibitem[{Sarojini et~al.(2023)Sarojini, Ruh, Joshy, Yan, Ivanov, Scotzniovsky, Fletcher, Orndorff, Sperry, Gandarillas et~al.}]{sarojini2023large}
Sarojini, D., Ruh, M.~L., Joshy, A.~J., Yan, J., Ivanov, A.~K., Scotzniovsky, L., Fletcher, A.~H., Orndorff, N.~C., Sperry, M., Gandarillas, V.~E., et~al., \enquote{Large-Scale Multidisciplinary Design Optimization of an eVTOL Aircraft using Comprehensive Analysis,} \emph{AIAA SCITECH 2023 Forum}, 2023, p. 0146.
\newblock \doi{10.2514/6.2023-0146}.

\bibitem[{Chauhan and Martins(2020)}]{chauhan2020tilt}
Chauhan, S.~S., and Martins, J.~R., \enquote{Tilt-wing eVTOL takeoff trajectory optimization,} \emph{Journal of aircraft}, Vol.~57, No.~1, 2020, pp. 93--112.

\bibitem[{Li and Lee(2024)}]{li2024analytic}
Li, S.~K., and Lee, S., \enquote{Analytic Prediction of Rotor Broadband Noise with Serrated Trailing Edges with Applications to Urban Air Mobility Aircraft,} \emph{Journal of the American Helicopter Society}, 2024.

\bibitem[{Gandarillas et~al.(2024)Gandarillas, Joshy, Sperry, Ivanov, and Hwang}]{gandarillas2022novel}
Gandarillas, V., Joshy, A.~J., Sperry, M.~Z., Ivanov, A.~K., and Hwang, J.~T., \enquote{A graph-based methodology for constructing computational models that automates adjoint-based sensitivity analysis,} \emph{Structural and Multidisciplinary Optimization}, Vol.~67, No.~5, 2024, p.~76.

\bibitem[{Sperry et~al.(2023)Sperry, Kondap, and Hwang}]{sperry2023automatic}
Sperry, M., Kondap, K., and Hwang, J.~T., \enquote{Automatic adjoint sensitivity analysis of models for large-scale multidisciplinary design optimization,} \emph{AIAA AVIATION 2023 Forum}, 2023, p. 3721.
\newblock \doi{10.2514/6.2023-3721}.

\bibitem[{Wang et~al.(2024{\natexlab{c}})Wang, Sperry, Gandarillas, and Hwang}]{wang2023accelerating}
Wang, B., Sperry, M., Gandarillas, V.~E., and Hwang, J.~T., \enquote{Accelerating model evaluations in uncertainty propagation on tensor grids using computational graph transformations,} \emph{Aerospace Science and Technology}, Vol. 145, 2024{\natexlab{c}}, p. 108843.
\newblock \doi{10.1016/j.ast.2023.108843}.

\bibitem[{Wang et~al.(2024{\natexlab{d}})Wang, Orndorff, and Hwang}]{wang2024partial}
Wang, B., Orndorff, N.~C., and Hwang, J.~T., \enquote{Graph-accelerated non-intrusive polynomial chaos expansion using partially tensor-structured quadrature rules for uncertainty quantification,} \emph{Aerospace Science and Technology}, Vol. 155, 2024{\natexlab{d}}, p. 109607.

\bibitem[{Katz(2001)}]{katz2001low}
Katz, J., \emph{Low-speed aerodynamics}, Cambridge University Press, 2001.

\bibitem[{Epton and Magnus(1990)}]{epton1990pan}
Epton, M.~A., and Magnus, A.~E., \enquote{PAN AIR: A computer program for predicting subsonic or supersonic linear potential flows about arbitrary configurations using a higher order panel method. Volume 1: Theory document (version 3.0),} Tech. rep., NASA, 1990.

\bibitem[{Hess and Smith(1967)}]{hess1967calculation}
Hess, J.~L., and Smith, A.~O., \enquote{Calculation of potential flow about arbitrary bodies,} \emph{Progress in Aerospace Sciences}, Vol.~8, 1967, pp. 1--138.

\bibitem[{Maskew(1987)}]{maskew1987program}
Maskew, B., \enquote{Program VSAERO theory Document: a computer program for calculating nonlinear aerodynamic characteristics of arbitrary configurations,} Tech. rep., NASA, 1987.

\bibitem[{Feinberg and Langtangen(2015)}]{feinberg2015chaospy}
Feinberg, J., and Langtangen, H.~P., \enquote{Chaospy: An open source tool for designing methods of uncertainty quantification,} \emph{Journal of Computational Science}, Vol.~11, 2015, pp. 46--57.
\newblock \doi{10.1016/j.jocs.2015.08.008}.

\bibitem[{Bouhlel et~al.(2019)Bouhlel, Hwang, Bartoli, Lafage, Morlier, and Martins}]{SMT2019}
Bouhlel, M.~A., Hwang, J.~T., Bartoli, N., Lafage, R., Morlier, J., and Martins, J. R. R.~A., \enquote{A Python surrogate modeling framework with derivatives,} \emph{Advances in Engineering Software}, 2019, p. 102662.
\newblock \doi{https://doi.org/10.1016/j.advengsoft.2019.03.005}.

\bibitem[{Rahman and Xu(2004)}]{rahman2004univariate}
Rahman, S., and Xu, H., \enquote{A univariate dimension-reduction method for multi-dimensional integration in stochastic mechanics,} \emph{Probabilistic Engineering Mechanics}, Vol.~19, No.~4, 2004, pp. 393--408.
\newblock \doi{10.1016/j.probengmech.2004.04.003}.

\end{thebibliography}
\end{document}